\documentclass{article}

\usepackage{graphicx}
\usepackage{txfonts}
\usepackage{subcaption}
\usepackage{arxiv}

\usepackage[utf8]{inputenc} 
\usepackage[T1]{fontenc}    
\usepackage{hyperref}       
\usepackage{url}            
\usepackage{booktabs}       
\usepackage{amsfonts}       
\usepackage{nicefrac}       
\usepackage{microtype}      
\usepackage{lipsum}		
\usepackage{graphicx}
\usepackage{natbib}
\usepackage{doi}

\title{Empirical measurement of cosmic luminosity-angular distance relation} 

\author{ \href{https://orcid.org/0000-0001-8318-6813}{\includegraphics[scale=0.06]{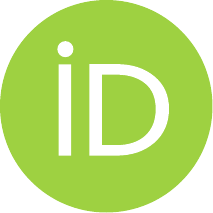}\hspace{1mm}Juan De Vicente}\\
	Centro de Investigaciones Energ\'eticas, Medioambientales y Tecnol\'ogicas (CIEMAT),\\Avda. Complutense 40, E-28040, Madrid, Spain\\
              \texttt{juan.vicente@ciemat.es} }

\renewcommand{\shorttitle}{Cosmic luminosity-angular distance relation}

\hypersetup{
pdftitle={Cosmic luminosity-angular distance relation},
pdfsubject={astrophysics, cosmology},
pdfauthor={Juan De Vicente},
pdfkeywords={cosmological parameters, distances and redshift},
}

\begin{document}
\maketitle

\begin{abstract}
	
One of the keys to understanding the universe is the proper measurement of  the relation between the luminosity distance $d_L$ and the angular diameter distance $d_A$. In 1933 Etherington deduced from general relativity the reciprocity equation $d_L=d_A(1+z)^{\gamma}$, with $\gamma=1$ for a local (non-expanding) universe. This relation has been adapted to an expanding universe with the value $\gamma=2$ by the introduction of the concept of comoving distance ($d_M$). In this work, we developed the method \textit{Cosmic Redshift Inference} to measure experimentally the value of $\gamma$ defined above independently of any possible galaxy evolution. The method has been applied to 1.2 million galaxy of SDSS DR15 obtaining the value $\gamma\simeq 1$.
	
\end{abstract}


\section{Introduction}

During the 20th century were established the foundations of modern cosmology. The field equations of general relativity were formulated by ~\citet{einstein1915feldgleichungen}. The definition of new metrics based on the cosmological principle with the properties of homogeneity and isotropy allowed the physicists the application of Einstein's field equations to the universe. While Einstein defined a static metric, ~\citet{friedmann1922krummung} deduces mathematically a non-stationary model with a time-dependent factor $a(t)$. The solution was independently derived by ~\citet{lemaitre1927lemaitre} interpreting a(t) as a scale factor of an expanding universe. The work was completed by ~\citet{robertson1933relativistic} and ~\citet{walker1937milne} in what is known as the \textit{Friedmann-Lema{\^\i}tre-Robertson-Walker} (FLRW) metric.  
	
	Contemporaneously to these achievements, a correlation between redshifts and distances for extragalactic sources was found by ~\citet{hubble1929relation}. The origin of this correlation was subject of intense debate between proponents of non-expanding and expanding universes on the 1930s (~\citet{kragh2017universe}). The fault of the Einstein's static universe to explain the redshift of galaxies leans the balance to the FLRW metric, whose time dependent factor $a(t)$ can directly explain the cosmological redshift (~\citet{tolman1934relativity}). The FLRW model describes a solutions to the Einstein's field  equations for a homogeneous and isotropic universe. The evolution and fate of the Universe depends on the nature of different density components, i.e., radiation, matter, curvature and dark energy. But as shown below, the FLRW metric also support a non-expanding universe accounting for the observed cosmological redshift.
	
	Different cosmological tests were proposed to probe whether the Universe is expanding or remains static. ~\citet{tolman1930estimation} predicted that in an expanding universe, the surface brightness of a receding source with redshift $z$ will be dimmed by $\sim(1+z)^{-4}$. Consequently to Tolman's prediction, the equation $d_L=d_A(1+z)^\gamma$, with $\gamma=2$ was established between \textit{luminosity distance} $d_L$ and \textit{angular diameter distance} $d_A$ for a expanding universe. In spite of some attempts (~\citet{,lubin2001tolman,sandage2010tolman}), these relations have not found a conclusive experimental support from cosmological surveys. On the other hand, the time dilation of Type Ia supernovae light curves suggested by ~\citet{wilson1939} and confirmed by ~\citet{goldhaber2001timescale} are assumed in favor of cosmological expansion though the same phenomenon can be described in a non-expanding universe as shown below.
	
	In this work, we have developed the method \textit{Cosmic Redshift Inference} (CRI) to determine experimentally the value of $\gamma$. The method has been applied to 1.2 million galaxies from SDSS DR15 sample (\citet{aguado2019fifteenth}), finding $\gamma \simeq 1$ which support a non-expanding universe rather than an expanding one. As ~\citet{tolman1934relativity} wrote "it is observations rather than hypothesis that must dictate the final nature of our cosmological theory".
Therefore, we have to find a non-expanding solution within the general relativity --different to the unsuccessful  Einstein's static universe-- to conciliate the theory and the experimental results. The most conservative approach is to look into the accepted FLRW metric. Note that this metric was first mathematically developed by ~\citet{friedmann1922krummung} and admits a different interpretation to the expanding one given by ~\citet{lemaitre1927lemaitre}. 

	The rest of the paper is organized as follows: In Section ~\ref{sec:stdModel} some basic distance definitions of the \textit{standard model} are reviewed. Section ~\ref{sec:lax} describes a new method to measure the relation between the luminosity distance and the angular diameter distance. Weakness of the current expanding FLRW model are shown in Section ~\ref{sec:eFLRW}. Section ~\ref{sec:sFLRW} shows the non-expanding interpretation of the FLRW metric. The conclusions are presented in Section ~\ref{sec:conclusions}. In Appendix ~\ref{sec:causesredshift}, a possible explanation of the redshift within a non-expanding universe is discussed.

  
 \section{Standard model of cosmology}
\label{sec:stdModel}

According to the standard model, the \textit{Friedmann-Lemaitre-Robertson-Walker} (FLRW) metric along with the Einstein's field equation of \textit{general relativity} describe a homogeneous and isotropic expanding universe. The FLRM metric is given by

\begin{eqnarray}
  \label{eq:flrw0}
   -c^2d\tau^2=-c^2dt^2+a(t)^2\left[\frac{dr^2}{1-kr^2}+r^2d\Omega^2\right]
\end{eqnarray}

\noindent
being

\begin{eqnarray}
  \label{eq:flrw_angles}
   d\Omega^2=d\theta^2+sin^2\theta d\phi^2
\end{eqnarray}

\noindent
where k describes the curvature while a(t) is the scale factor accounting for the universe expansion. There are different distance ladders relating theory and observations. Let us to provide a brief summary of some distance definitions and their relations with normalized densities ($\Omega_M$, $\Omega_r$, $\Omega_\Lambda$, $\Omega_k$), corresponding to matter, radiation, cosmological constant and curvature (\cite{hogg1999distance}). The first Friedmann equation can be expressed from the Hubble parameter $H$ at any time, and the Hubble constant $H_0$ today as

\begin{eqnarray}
  \label{eq:friedmann}
   \frac{\dot{a}(t)^2}{a(t)^2}=H^2=H_0^2 E(z)^2
\end{eqnarray}

\noindent
where 

\begin{eqnarray}
\label{eq:ez}
  E(z)=\sqrt{\Omega_K(1+z)^2+\Omega_\Lambda+\Omega_M(1+z)^3+\Omega_r(1+z)^4}
\end{eqnarray}

\noindent
By integrating Eq. ~\ref{eq:flrw0} along with Eq. ~\ref{eq:friedmann} one can obtain the line of sight \textit{comoving distance} $d_C$ as

\begin{eqnarray}
\label{eq:comovingDistance}
  d_C=d_H\int_{0}^{z}\frac{dz'}{E(z')}
\end{eqnarray}

\noindent
where $d_H=c/H_0=3000 h^{-1} Mpc$ is the \textit{Hubble distance}. From the same equations one can get the \textit{transverse comoving distance} $d_M$ as

\begin{eqnarray}
\label{eq:transverseComovingDistance}
d_M= 
\left \{ 
 \begin{array}{ccc}
    d_H\frac{1}{\sqrt\Omega_k}\sinh[\sqrt\Omega_kd_C/d_H]&for&\Omega_k>0 \\ 
    d_c&for&\Omega_k=0 \\
    d_H\frac{1}{\sqrt{|\Omega_k|}}\sin[\sqrt{|\Omega_k|}d_C/d_H]&for&\Omega_k<0 \\
    \end{array}
\right \}
\end{eqnarray}

\begin{figure}[b]
    \centering
    \leavevmode
      \includegraphics[width=0.5\textwidth]{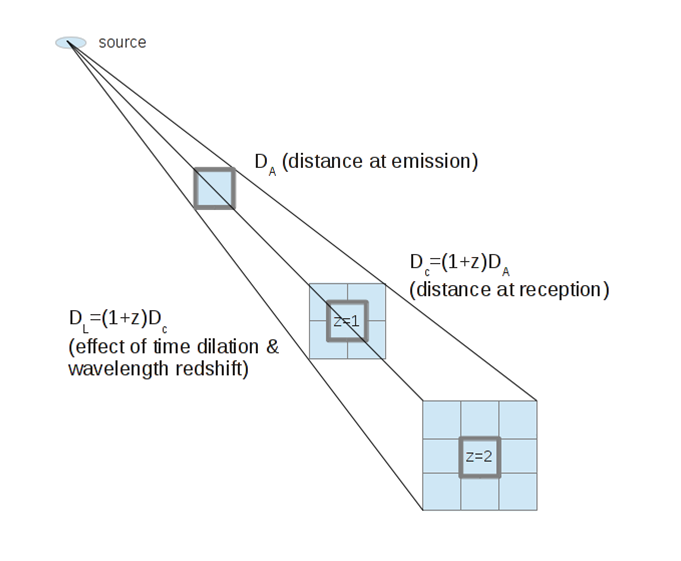}
			\captionsetup{labelfont=bf}
      \caption{\textit{Standard Model luminosity-angular distances relation}: \textit{Angular diameter distance} ($d_A$), \textit{comoving distance} ($d_C$) and \textit{luminosity distance} ($d_L$) for a flat universe. $d_A$ is the distance at emission, $d_C$ is the distance at reception and $d_L$ account for the distance elongation due to universe expansion ($\sim(1+z)$), time dilation and wavelength redshifting ($\sim(1+z))$. The relation $d_L=d_A(1+z)^2$ can be deduced from the figure.
       \label{fig:gauss3D_short}}
 \end{figure}%

\noindent
With respect to observable quantities, the \textit{angular diameter distance} $d_A$ is defined as the ratio between the object physical size $S$ and its angular size $\theta$ 
 
\begin{eqnarray}
\label{eq:angularDiameterDistanceTh}
  d_A=\frac{S}{\theta}
\end{eqnarray}

\noindent
The \textit{angular diameter distance} is related to the \textit{transverse comoving distance} by
 
\begin{eqnarray}
\label{eq:angular_vs_transverseComoving}
  d_M=d_A(1+z)
\end{eqnarray}

\noindent
where z is the redshift. On the other hand, the \textit{luminosity distance} defines the relation between the bolometric flux energy $f$ received at earth from an object, to its bolometric luminosity L by means of

\begin{eqnarray}
\label{eq:fluxEnergy}
 f= \frac{L}{4\pi d_L^2} 
\end{eqnarray}

\noindent
or finding $d_L$

\begin{eqnarray}
\label{eq:DL_th}
 d_L=\sqrt{\frac{L}{4\pi f}} 
\end{eqnarray}

\noindent
The relation between $d_L$ and $d_M$ is given by

\begin{eqnarray}
\label{eq:luminosity_vs_transverseComoving}
d_L= d_M(1+z)
\end{eqnarray}

\noindent
and taking into account Eq. ~\ref{eq:angular_vs_transverseComoving}

\begin{eqnarray}
\label{eq:luminosity2_vs_angular2}
d_L^2= d_A^2(1+z)^4
\end{eqnarray}

\noindent
There are four (1+z) factors affecting to flux energy diminution (Fig.~\ref{fig:gauss3D_short}). Two come from the elongation of the initial distance $d_A$ by a factor of $(1+z)$ due to universe expansion according to the inverse square law. Another factor comes from the time dilation due to universe expansion that reduces the photon emission/reception rate by $(1+z)^{-1}$. The last factor comes from the cosmological wavelength redshift that decrease the energy of photons by $(1+z)^{-1}$. Therefore, a relevant relation is established between the \textit{angular diameter distance} and the \textit{luminosity distance} in the \textit{expanding universe} as

\begin{eqnarray}
\label{eq:stDistancesRelation}
d_L=d_A(1+z)^2               \qquad (expanding\ universe)
\end{eqnarray}

\noindent
Eq. ~\ref{eq:stDistancesRelation} is commonly known as Etherington distance-duality relation.

\section{Cosmic Redshift Inference: Empirical determination of the luminosity-angular distances relation}
\label{sec:lax}

In this section we derive a photometric redshift method based on the cosmological luminosity-angular distances relation.

The luminosity distance can be expressed as

\begin{eqnarray}
\label{eq:dl}
    d_L=\sqrt{\frac{L}{4\pi f_L}}
\end{eqnarray}

\noindent
and since the angular distance corresponds to the distance at emission for both non-expanding and expanding universes, it can be expressed also as

\begin{eqnarray}
\label{eq:da}
	d_A=\sqrt{\frac{L}{4\pi f_A}}
\end{eqnarray}

\noindent
where $f_A$ corresponds to the hypothetical flux that would be measured at a distance $d_A$ without neither expansion nor redshift. Dividing both expressions we have

\begin{eqnarray}
\label{eq:dlda}
\frac{d_L}{d_A}=\sqrt{\frac{f_A}{f_L}}
\end{eqnarray}

On the other hand, the luminosity-angular distances relation is given by

\begin{eqnarray}
\label{eq:ci}
    (1+z)^{\gamma}=\frac{d_L}{d_A}
\end{eqnarray}

\noindent
with $\gamma=2$ for an expanding universe and $\gamma=1$ for a non-expanding one. Substituting Eq.~\ref{eq:dlda} in Eq.~\ref{eq:ci} we have
 
 \begin{eqnarray}
\label{eq:lax}
    (1+z)^{\gamma}=\left[\frac{f_A}{f_L}\right]^{1/2}
\end{eqnarray}

\noindent Taking base10 logarithm and multiplying by 2.5 in both sides of the equation we have

\begin{eqnarray}
\label{eq:lax1}
    5 \gamma \log(1+z)=-2.5log{f_L}-(-2.5log{f_A})
\end{eqnarray}

\noindent and defining \\
\\
$m_L=-2.5\log{f_L}$ as the luminosity magnitude \\
$m_A=-2.5\log{f_A}$ as the angular magnitude \\
$m_{\gamma}=5 \gamma \log(1+z)$ as the redshift magnitude \\

\noindent we have
\begin{eqnarray}
\label{eq:lax2}
    m_L = m_A + m_{\gamma}
\end{eqnarray}

The equation can also be expressed for common multi-band surveys in a vectorial form as

\begin{eqnarray}
\label{eq:lax3}
   \mathbf { m_L = m_A + m_{\gamma} }                                   
\end{eqnarray}

Note that the luminosity magnitude has two independent components: $\mathbf{m_A}$ that depends on the luminosity of the source and the distance at emission, and $\mathbf{m_\gamma}$ that depends exclusively on redshift. Multiplying Eq.~\ref{eq:lax3} by an unitary vector $\mathbf{v_\gamma}$ in the direction of $\mathbf{m_\gamma}$, we obtain

\begin{eqnarray}
\label{eq:lax4}
   \mathbf { m_L \cdot v_\gamma = m_A \cdot v_\gamma + m_{\gamma} \cdot v_\gamma}                       
\end{eqnarray}

As $\mathbf {m_{\gamma} \cdot v_\gamma}=m_\gamma$ and $\mathbf {m_A \cdot v_\gamma}=0$ since both vectors are orthogonal, the expression becomes   

\begin{eqnarray}
\label{eq:lax5}
   m_{\gamma}=\mathbf {m_L \cdot v_\gamma}                                   
\end{eqnarray}

To assess the validity of this expression, we need a galaxy sample with spectroscopic redshift ($m_{\gamma}$) and properly measured photometric magnitudes for an arbitrary number of bands ($\mathbf {m_L}$). A regression can be applied to this sample to determine $\mathbf {v_{\gamma}}$. The success of this approximation requires a high correlation between both sides of Eq.~\ref{eq:lax5}. The proper value of $\gamma$ is the one whose corresponding computed $\mathbf{v_{\gamma}}$ is unitary as assumed above ($\|\mathbf{v_\gamma}\|=1$).

\begin{figure} 
  \centering
         \includegraphics[width=0.53\textwidth]{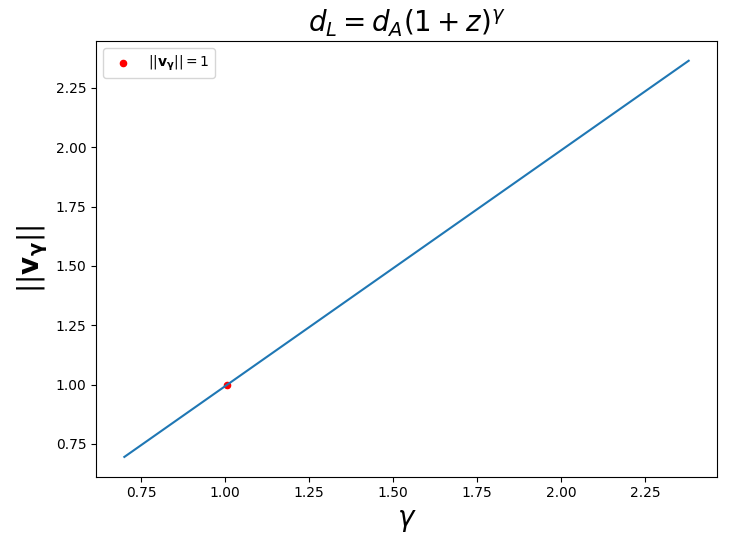}
  \captionsetup{labelfont=bf} \captionof{figure}{Cosmic Redshift Inference: A linear regression is performed to measure the value of $\gamma$. The redshift vector should meet $||v_\gamma||=1$. It occurs for $\gamma\simeq 1$}
 \label{fig:gammaDeterminationVnorm}
\end{figure} 

\begin{figure*}
  \centering
  \begin{tabular}{c @{\qquad} c }
    \includegraphics[width=.48\linewidth]{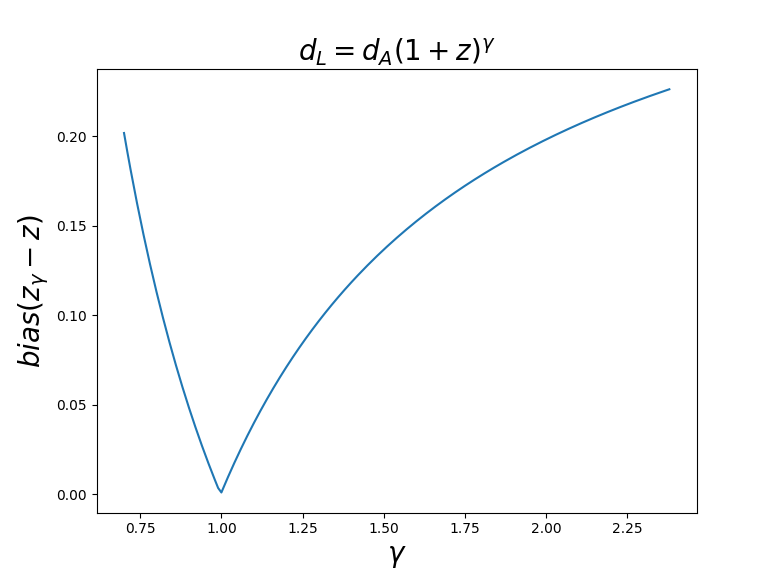} &
    \includegraphics[width=.48\linewidth]{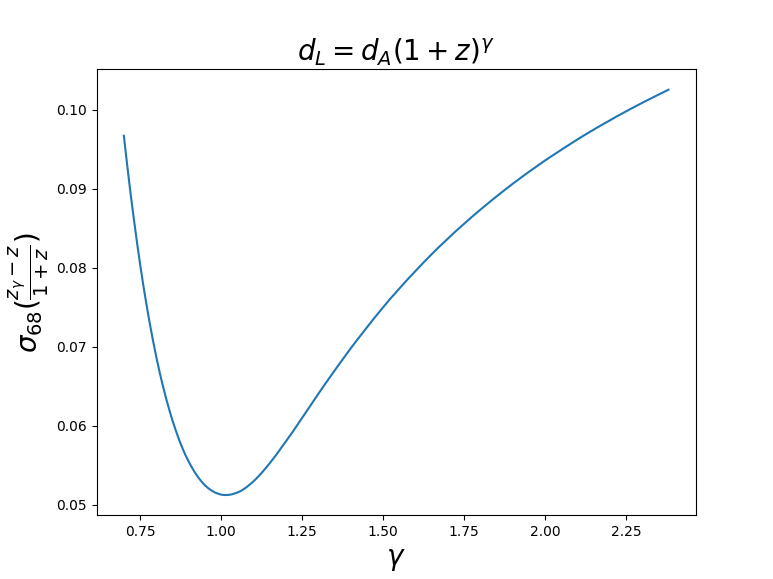} \\
    \small (a)  & \small (b)
  \end{tabular}
	\captionsetup{labelfont=bf}
  \caption{Cosmic Redshift Inference: Redshift reconstruction after applying $||v_\gamma||=1$: (a) bias of $\delta(z-z_\gamma)$ distribution. (b) $\sigma_{68}$ of $\delta(z-z_\gamma)/(1+z)$ distribution}
\label{fig:gammaDeterminationbiass68}
\end{figure*}

\begin{figure*}
  \centering
  \begin{tabular}{c @{\qquad} c }
    \includegraphics[width=.48\linewidth]{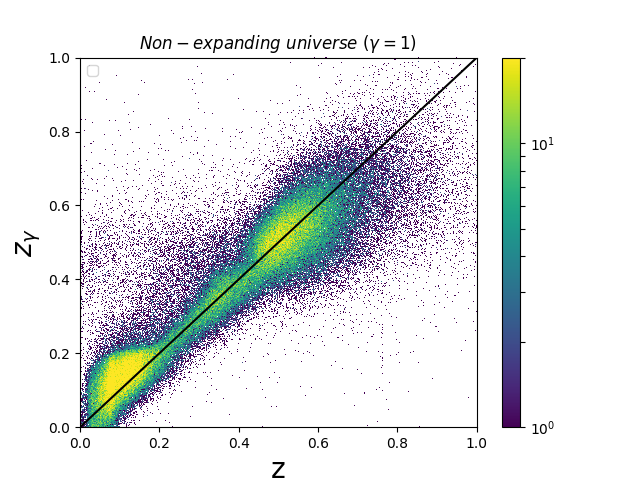} &
    \includegraphics[width=.48\linewidth]{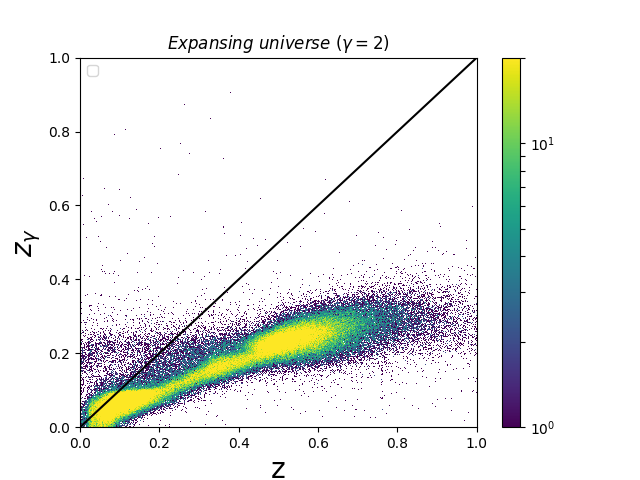} \\
    \small (a)  & \small (b)
  \end{tabular}
	\captionsetup{labelfont=bf}
	\captionof{figure}{Scatter plot from the redshift reconstruction after applying $||v_\gamma||=1$: (a) Non-expanding universe (b) Expanding universe}
  \label{fig:static_expanding}
\end{figure*} 

We have resorted to SDSS DR15 that provides simultaneously spectra and photometric measurements for about 1.2 million galaxies. To ensure a uniform treatment for all galaxies, we select \textit{De Vaucouleurs} magnitudes ($deVMag_{ugriz}$) which achieves accurate measurement of the flux of the bulge, the most luminous part of the galaxies. Thus, $\mathbf{m}=(deVMag_u,deVMag_g,deVMag_r,deVMag_i,deVMag_z,1)$, where the last component will account in the regression for the unmatched zero definition between redshift and magnitudes.


We have applied the regression (Eq.~\ref{eq:lax5}) to this sample obtaining a parameter vector $\mathbf{v_{\gamma}}$ that provides high correlation ($c=0.91$) between both sides of the equation for any value of $\gamma$. Nevertheless, only one value of $\gamma$ meets the second condition: $\|\mathbf{v_\gamma}\|=1$ (Fig.~\ref{fig:gammaDeterminationVnorm}). We can see that this value corresponds to $\gamma\simeq 1$ which is not expected for an expanding universe but for a non-expanding one.

Another way to determine the proper value of $\gamma$ is to inversely reconstruct the value of the redshift $z_\gamma$ after normalizing the measured value of $\mathbf{v_{\gamma}}$. Finding $z$ from the definition of $m_\gamma$ and taking into account Eq.~\ref{eq:lax5} we have



\begin{eqnarray}
\label{eq:lax9}
    z_\gamma=10^\frac{\mathbf {m_L \cdot v_{\gamma}}}{5\gamma}-1
\end{eqnarray}

In such a way that we can compare $z_\gamma$ with the spectroscopic value $z$. Fig.~\ref{fig:gammaDeterminationbiass68} shows the results of redshift reconstruction along a range of $\gamma$ values. The Fig.~\ref{fig:gammaDeterminationbiass68}a corresponds to the bias for $\delta z=(z-z_\gamma)$ distribution which shows a minimum at $\gamma\simeq 1$ which corresponds to a non-expanding universe. In Fig.~\ref{fig:gammaDeterminationbiass68}b the ordinate axis represents $\sigma_{68}$ that is a measurement of the dispersion of the distribution $\delta z=(z-z_\gamma)/(1+z)$. It corresponds to the width of the distribution measured with respect to the median, in which $68\%$ of the galaxies are enclosed.

\begin{eqnarray}
\label{eq:lax10}
    \sigma_{68}=\frac{1}{2}(P_{84}-P_{16})
\end{eqnarray}

where $P_{16}$ and $P_{84}$ are the $84th$ and the $16th$ percentile of the cumulative distribution respectively. The abscissa axis corresponds to $\gamma$. The minimum deviation is also at $\gamma\simeq 1$.

Fig.~\ref{fig:static_expanding} shows the scatter plot of redshift reconstruction for $\gamma=1$ \textit{(non-expanding universe)} and $\gamma=2$ \textit{(expanding universe)}. The comparison support a non-expanding universe.

\section{Weakness of expansion interpretation of FLRW metric}
\label{sec:eFLRW}

At this point we can rise the question: \\
What are the equations that support the expanding universe? 
The expanding universe rest on the metric tensor $g_{\mu\nu}$ for a homogeneous and isotropic universe given by the FLRW metric and on the Einstein's field equation.

\begin{eqnarray}
\label{eq:EFE}
	R_{\mu\nu}-\frac{1}{2}g_{\mu\nu}R+\Lambda g_{\mu\nu}=\frac{8 \pi G}{c^4} T_{\mu\nu}
\end{eqnarray}

But that is not all, the expanding universe requires the additional support from the equation 

\begin{eqnarray}
\label{eq:dLdM}
	d_L=d_M(1+z)
\end{eqnarray}

which relates the luminosity distance $d_L$ with the transverse comoving distance $d_M$, equation not derived from Einstein's Field equations in spite of both the energy-momentum tensor $T_{\mu\nu}$ and the luminosity distance $d_L$ are related to energy fluxes. Note that it is not $a(t)$ but the definition of this comoving distance $d_M$ and the consideration of $r$ in FLRW metric as a comoving coordinate that force the FLRW model to be an expanding one.

\begin{figure}[b]
    \centering
    \leavevmode
      \includegraphics[width=0.50\textwidth]{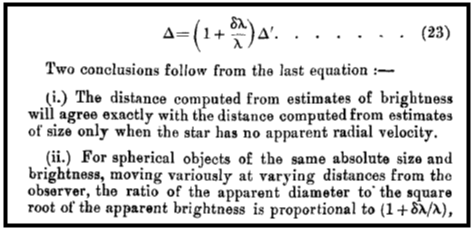}            
			\captionsetup{labelfont=bf}
      \caption{Etherington equation (snapshot from his paper).}
       \label{fig:EtheringtonSnapshot}
 \end{figure}%

A similar question arise with respect to Etherington relation. Fig.~\ref{fig:EtheringtonSnapshot} shows the relation between the luminosity distance $\Delta$ and the angular diameter distance $\Delta'$ derived by Etherington. Thus, the Etherington equation deduced from general relativity was the reciprocity theorem for a non-expanding universe:

 \begin{eqnarray}
\label{eq:etheringtonEquation}
	d_L=d_A(1+z)			\qquad (non-expanding\ universe)
\end{eqnarray}

\noindent Note that this equation can be adapted for an expanding universe by the introduction of an additional intermediate variable $d_M$ (comoving distance) --not considered by Etherington--, and taking into account Eq.~\ref{eq:dLdM} that combined with Eq.~\ref{eq:dMdA}

\begin{eqnarray}
\label{eq:dMdA}
	d_M=d_A(1+z)
\end{eqnarray}

gives

\begin{eqnarray}
\label{eq:dLdA}
	d_L=d_A(1+z)^2			\qquad (expanding\ universe)
\end{eqnarray}

\noindent equation known as Etherington duality relation.

Therefore, it is by the definition of these two external equations (Eq.~\ref{eq:dLdM} and Eq. ~\ref{eq:dMdA})) --non derived from general relativity-- that the expanding paradigm is supported.

\section{Non-expanding interpretation of FLRW metric}
\label{sec:sFLRW}

Let us to consider the FLRW metric given by

\begin{eqnarray}
  \label{eq:flrw}
   -c^2d\tau^2=-c^2dt^2+a(t)^2\left[\frac{dr^2}{1-kr^2}+r^2d\Omega^2\right]
\end{eqnarray}

\begin{figure}[b]
    \centering
    \leavevmode
      \includegraphics[width=0.5\textwidth]{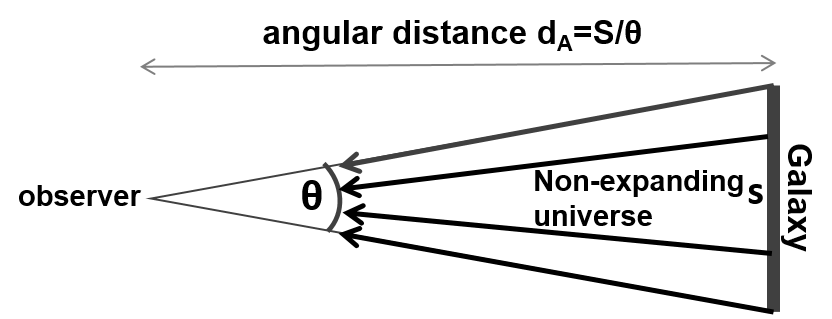}
			\captionsetup{labelfont=bf}
      \caption{\textit{Geodesics for light rays in a flat non-expanding FLRW universe.}}
       \label{fig:nonexpanding_universe}
 \end{figure}%

Let us to analyze the behavior of light rays in FLRW metric considering a non-expanding universe (Fig.~\ref{fig:nonexpanding_universe}). Let the origin of coordinates be at the observer O, and considers an extended cosmological object (galaxy) initially located at a distance $d_A$ from O at time of emission $t_e$. According to general relativity, light rays follow null geodesics where $d\tau^2=0$. Substituting this value in Eq.~\ref{eq:flrw}, light rays follow the equation

\begin{eqnarray}
  \label{eq:flrw_ct}
   c^2dt^2=a(t)^2\left[\frac{dr^2}{1-kr^2}+r^2d\Omega^2\right] 
\end{eqnarray}

Note from Fig.~\ref{fig:nonexpanding_universe} that in a non-expanding universe the light rays that will arrive to the observer O in the future are those pointing initially towards the observer at time of emission $t_e$ (leaving apart non pure cosmological effects as gravitational lensing or astrophysical events). These rays will maintain the same direction up to arriving to the observer, i.e., $\theta=cte$, $d\theta=0$, $\phi=cte$, $d\phi=0$ in Eq.~\ref{eq:flrw_angles}, and hence $d\Omega=0$ . Substituting $d\Omega=0$ in Eq.~\ref{eq:flrw_ct}, the light rays that will arrive to the observer meet

\begin{eqnarray}
  \label{eq:flrw_radial}
   c^2dt^2=a(t)^2\frac{dr^2}{1-kr^2} 
\end{eqnarray}

Integrating Eq.~\ref{eq:flrw_radial} from $t_e=0$ we have

\begin{eqnarray}
  \label{eq:flrw_radial_int}
   \int_0^t{\frac{cdt}{a(t)}}=\int_0^{d_M}{\frac{dr}{\sqrt{1-kr^2}}} 
\end{eqnarray}

\noindent
Note that in a non-expanding universe, the comoving coordinate $r$ of the FLRW model looses its meaning and r can be interpreted as luminosity (distance) coordinate. In this way, the comoving distance $d_M$ can be substituted by the luminosity distance $d_L$ in all equation. Therefore

\begin{eqnarray}
\label{eq:ExpansionLensingDL}
  d_L=d_M 			\qquad (non-expanding\ universe)
\end{eqnarray} 

and 

\begin{eqnarray}
  \label{eq:flrw_radial_int_L}
   \int_0^t{\frac{cdt}{a(t)}}=\int_0^{d_L}{\frac{dr}{\sqrt{1-kr^2}}} 
\end{eqnarray}

Thus, in the case of a flat ($k=0$) non-expanding universe, we would have

\begin{eqnarray}
  \label{eq:flrw_radial_int_LK}
   d_L=\int_0^t{\frac{cdt}{a(t)}} 
\end{eqnarray}

which would be the factor responsible of time dilation and cosmological redshift in a non-expanding universe. For the case where c is constant, we can define 

\begin{eqnarray}
\label{eq:deltat}
\Delta t'=\int_0^t{\frac{dt}{a(t)}}
\end{eqnarray}

as time dilation in such a way that $d_A=c \Delta t$ and $d_L=c \Delta t'$.\\

On the other hand, Eq.~\ref{eq:transverseComovingDistance} transforms to Eq.~\ref{eq:DLtransverseComovingDistance}

\begin{eqnarray}
\label{eq:DLtransverseComovingDistance}
d_L= 
\left \{ 
 \begin{array}{ccc}
    d_H\frac{1}{\sqrt\Omega_k}\sinh[\sqrt\Omega_kd_C/d_H]&for&\Omega_k>0 \\ 
    d_c&for&\Omega_k=0 \\
    d_H\frac{1}{\sqrt{|\Omega_k|}}\sin[\sqrt{|\Omega_k|}d_C/d_H]&for&\Omega_k<0 \\
    \end{array}
\right \}
\end{eqnarray}

that directly relates the first Friedmann equation with the observable $d_L$ derived from FLRW metric (Eq.~\ref{eq:flrw_radial_int_LK}) without the need to define additional equations detached from general relativity. Let us to express the FLRW metric in another form. Dividing Eq.~\ref{eq:flrw} by $a(t)^2$ we have Eq.~\ref{eq:flrw_a}. Note that in this form it is more clear the time dilation nature of $a(t)$ rather than the role of a scale factor for an expanding universe.

\begin{eqnarray}
  \label{eq:flrw_a}
   -\left(\frac{c d\tau}{a(t)}\right)^2=-\left(\frac{c dt}{a(t)}\right)^2+\left[\frac{dr^2}{1-kr^2}+r^2d\Omega^2\right]
\end{eqnarray}


In a expanding universe, the first Friedmann equation constraints the form of the scale factor $a(t)$ with the different species of the universe as radiation, matter, curvature and cosmological constant. According to the present experimental results and the non-expanding interpretation of the FLRW model, $a(t)$ corresponds to another unidentified property of vacuum --different from expansion-- which also depends on the relative content of these species along cosmic time. 

The stability of the non-expanding universe resides on the acceptance of the cosmological principle since $a(t)$ is no more related to the size of the universe. The standard model demonstrates that the geodesics for a free particle (galaxy) corresponds to fixed FLRW comoving coordinates. The same demonstration applies to non-expanding-FLRW by interpreting comoving coordinates as luminosity (distance) ones. Thus, geodesics for a free particle in a non-expanding universe corresponds to fixed FLRW luminosity coordinates, which are spatially fixed but affected by a time dilation in the reception of signals. Therefore, it is the own FLRW metric that ensure the stability of the non-expanding universe.

\section{Conclusions}
\label{sec:conclusions}

In the 1930s, early after the discovery of the redshift-distance relation, a debate emerged among physicist regarding the feasibility of a non-expanding or an expanding universe. Tolman proposed a surface brightness test as a mean to differentiate an expanding from a non-expanding universe. The test predicts the relation $\mu\sim(1+z)^{-4}$ for an expanding universe. It corresponds to $d_L=d_A(1+z)^\gamma$, with $\gamma=2$ for an expanding universe and $\gamma=1$ for a non-expanding one. Up to now, non-conclusive test have been performed on galaxy survey in this respect due to the unknown possible galaxies' evolution. 

	In this work we have developed Cosmic Redshift Inference, a photometric redshift method that it is able to isolate the dimming due to the redshift from the measured magnitude, allowing us to measure  the value of $\gamma$. The method has been applied to 1.2 million galaxy of the public SDSS DR15 catalog. The result obtained $\gamma\simeq 1$ supports a non-expanding universe. 
	
	The expanding universe is supported by the Einstein's Field Equations and the FLRW metric. In this work, we note that the expanding interpretation of the FLRW metric also requires some external equation based on the definition of the comoving concept. Striping away this supplement from the FLRW model, uncover the non-expanding essence of the metric while still preserves $a(t)$, the time dependent function responsible of the redshift. Thus, the non-expanding universe determined experimentally can still be accommodated within the FLRW model. It only requires the interpretation of the ad hoc comoving coordinates as the observed luminous ones. 
	
	The research is not over and the physical cause of the cosmological redshift have to be determined. In Appendix ~\ref{sec:causesredshift}, we explore the magnetic permeability variation with cosmic time as a possible cause of redshift. The growth of the magnetic permeability would drive to a decrease of speed of light while preserving the observed gross atomic structure of distant galaxies.  

 
\section*{Acknowledgements}
Funding support for this work was provided by the Autonomous Community 
of Madrid through the project TEC2SPACE-CM (S2018/NMT-4291).
 
This paper uses data from public SDSS DR-15. 
Funding for the Sloan Digital Sky Survey IV has been provided by the Alfred P. Sloan Foundation, the U.S. Department of Energy Office of Science, and the Participating Institutions. SDSS-IV acknowledges
support and resources from the Center for High-Performance Computing at
the University of Utah. The SDSS web site is www.sdss.org.

SDSS-IV is managed by the Astrophysical Research Consortium for the 
Participating Institutions of the SDSS Collaboration including the 
Brazilian Participation Group, the Carnegie Institution for Science, 
Carnegie Mellon University, the Chilean Participation Group, the French Participation Group, Harvard-Smithsonian Center for Astrophysics, 
Instituto de Astrof\'isica de Canarias, The Johns Hopkins University, Kavli Institute for the Physics and Mathematics of the Universe (IPMU) / 
University of Tokyo, the Korean Participation Group, Lawrence Berkeley National Laboratory, 
Leibniz Institut f\"ur Astrophysik Potsdam (AIP),  
Max-Planck-Institut f\"ur Astronomie (MPIA Heidelberg), 
Max-Planck-Institut f\"ur Astrophysik (MPA Garching), 
Max-Planck-Institut f\"ur Extraterrestrische Physik (MPE), 
National Astronomical Observatories of China, New Mexico State University, 
New York University, University of Notre Dame, 
Observat\'ario Nacional / MCTI, The Ohio State University, 
Pennsylvania State University, Shanghai Astronomical Observatory, 
United Kingdom Participation Group,
Universidad Nacional Aut\'onoma de M\'exico, University of Arizona, 
University of Colorado Boulder, University of Oxford, University of Portsmouth, 
University of Utah, University of Virginia, University of Washington, University of Wisconsin, 
Vanderbilt University, and Yale University.


\bibliographystyle{unsrtnat}
\bibliography{el2}  

\appendix
\label{appendix}

\section{Exploring variable magnetic permeability with cosmic time as cause of cosmological redshift}
\label{sec:causesredshift}

The physical causes of the redshift have to be identified among the different proposals. The most popular alternative approach to expansion as the cause of the redshift is the tired light (~\citet{zwicky1929possibilities}), where photons lose energy while traveling to earth. Another possibility to be explored is the variable speed of light (VSL) with cosmic time. Let us to extend about this possibility in this section. The cosmological principle assumes a universe spatially homogeneous and spatially isotropic. It does not state that the universe is the same over time. Thus, according to the cosmological principle we can allow a space property to change overtime. That is the case of the scale factor in the expanding universe or the speed of light in the non-expanding one. There are different VSL theories as those addressing the horizon problem (~\citet{moffat1993superluminary}, ~\citet{albrecht1999time}) or the ones allowing the variation of speed of light between free-falling observers (~\citet{dicke1957gravitation}). Other depart from FLRW metric as the one that assumes both expansion and VLS (~\citet{van2021general})(which would requires a value of $\gamma\geq3$, far from our measurement) or those assuming photons emitted at higher  speed of light at earlier times, but maintaining such high velocities up to earth (~\citet{pipino2021variable}), events not observed experimentally.\\

Though less known, there is a plausible alternative explanation to redshift based on variable speed of light (VSL) with cosmic time (~\citet{wold1935redward}). Such approach would still require a spatially constant speed of light among all free falling observes as the general relativity demands. In this case, from Eq. ~\ref{eq:flrw_radial_int_LK} we can write

\begin{eqnarray}
  \label{eq:flrw_radial_int_LL}
   d_L=\int_0^t{c(t)dt}=c_0\int_0^t{\frac{dt}{a(t)}}  
\end{eqnarray}

being $c_0$ the current value of the speed of light. 

The process of photon redshift based on a speed of light decreasing with cosmic time can be described as follow: A galaxy emits a photon at speed $c_z$ due to an electron transition between atomic levels at its corresponding energy $h\nu_0$, being lambda stretched out at emission due to the equation $\lambda_z=c_z/\nu_0$. In the travel of the photon to earth, $\lambda_z$ remains constant, while the frequency decrease up to $\nu_z$ due to speed of light drop $\nu_z=c_0/\lambda_z$. \\

Note that Eq.~\ref{eq:flrw_radial_int_LL} implies $d_L=d_A$, and the assumed Eq. ~\ref{eq:ci} changes to
 
\begin{eqnarray}
\label{eq:d0}
    (1+z)^{\gamma}=\frac{d_L}{d_0}
\end{eqnarray}

being $d_0$ the angular distance $d_A$ that we would have if the speed of light were constant along cosmic time.\\

Given that the speed of light is 

\begin{eqnarray}
  \label{eq:c_ep_mu}
   c=\frac{1}{\sqrt{\epsilon_0 \mu_0}} 
\end{eqnarray}

some of the vacuum properties either dielectric permittivity $\epsilon_0$ or magnetic permeability $\mu_0$ have to change with cosmic time. Since the gross atomic structure of redshifted galaxies and its corresponding energy levels depend on $\epsilon_0$, we assume that it is $\mu_0=\mu_0(t)$ that depends on cosmic time in the form

 \begin{eqnarray}
  \label{eq:flrw_mu0_a}
   a(t)=\sqrt{\mu_0(t)}
\end{eqnarray}

in such a way that only magnetic fields and the fine structure constant would be affected along cosmic time.
Thus, the time dependent function $a(t)$ defined in $FLRW$ metric would not correspond to an expansion but to a known property of vacuum, the square root of the magnetic permeability. Consequently, the speed of light would decreases with cosmic time $c(t)=1/\sqrt{\epsilon_0 \mu_0(t)}$ while the electric permittivity $\epsilon_0$ remains constant allowing the observed atomic structures. The model can be denominated $\mu_0-FLRW$ to differentiate it from other possible alternatives.

Note that $\mu_0-FLRW$ universe does not change the FLRW equations but reinterpreted them. Thus, $\mu_0-FLRW$ may assume the main ideas of the standard model as that the early universe was hotter and dominated by radiation, but the drop in universe temperature would not be due to expansion but to the drop in the speed of light with cosmic time. Thus, the cosmic microwave background would have been emitted at the same energy as in the standard model with values $\nu_z=\nu_{cmb}(1+zcmb)$, $c_z=c_0(1+zcmb)$ and $\lambda_{cmb}=c_z/\nu_z$, and is received as $c_0$, $\nu_{cmb}$ and $\lambda_{cmb}$.

\end{document}